\documentclass[useAMS,usenatbib]{mn2e}
\usepackage[percent]{overpic}
\usepackage{times}
\usepackage{graphicx}
\usepackage{amsmath}
\usepackage{mathabx}
\usepackage{subfigure}
\usepackage{natbib}
\usepackage{txfonts}
\usepackage{journals}
\usepackage{xcolor}
\usepackage{array}

\newcommand{\egk}{$E(G-K_{\rm S})$}
\newcommand{\ebr}{$E(G_{\rm BP}-G_{\rm RP})$}
 \newcommand{\ehk}{$E(H-K_{\rm S})$}
\newcommand{\ppi}{Paper~I}

\title[Distances of molecular clouds in disk]
{A large catalogue of molecular clouds with accurate distances within 4\,kpc of the Galactic disk}

\author[B.Q. Chen et al.]
{B.-Q. Chen,$^{1}$\thanks{E-mail:
bchen@ynu.edu.cn (BQC); gxli@ynu.edu.cn (GXL); x.liu@ynu.edu.cn (XWL).}
 G.-X. Li,$^1$\footnotemark[1]
 H.-B. Yuan,$^2$
 Y. Huang,$^{1}$
 Z.-J. Tian,$^{3}$
 H.-F. Wang,$^{1,4}$\thanks{LAMOST Fellow}
     \newauthor
 H.-W. Zhang,$^{5}$
C. Wang,$^{5}$\footnotemark[2]
and  X.-W. Liu$^{1}$\footnotemark[1]
\\
$^{1}$South-Western Institute for Astronomy Research, Yunnan University, Chenggong District, Kunming, 650500, P.\,R.\,China\\
$^{2}$Department of Astronomy, Beijing Normal University, Beijing 100875, P.\,R.\,China\\
$^{3}$Department of Astronomy, Yunnan University, Chenggong District, Kunming, 650500, P.\,R.\,China\\
$^{4}$Department of Astronomy, China West Normal University, Nanchong 637009, P.\,R.\,China\\
$^{5}$Department of Astronomy, Peking University, Beijing 100871, P.\,R.\,China\\
 }

\begin{document}

\date{Accepted ???. Received ???; in original form ???}

\pagerange{\pageref{firstpage}--\pageref{lastpage}} \pubyear{2016}
\maketitle
\label{firstpage}

\begin{abstract}
  We present a large, homogeneous catalogue of  molecular clouds within 4\,kpc from the Sun at
  low Galactic latitudes ($|b|$ $<$ 10\degr) with unprecedented accurate
  distance determinations.  Based on the three-dimensional
  dust reddening map and estimates of colour excesses and distances of over 32 million stars
  presented in Chen et al, we have identified 567 dust/molecular clouds with a hierarchical structure identification method
  and obtained their distance estimates by a dust model fitting algorithm.
  The typical distance uncertainty is less than 5\,per\,cent.
  As far as we know, this is the first large catalogue of molecular clouds in the Galactic plane  with
distances derived in a direct manner.
  The clouds are seen to lie along the Sagittarius,
  Local and Perseus Arms. In addition to the
  known structures, we propose the existence of a possible {\it spur}, with a pitch angle of about 34\degr,  connecting the Local
  and the Sagittarius Arms in the fourth quadrant. 
  We have also derived the physical properties of those molecular clouds.
  The distribution of cloud properties in different parameter spaces agrees grossly with the previous results.
  Our cloud sample is an ideal starting point to study the concentration of dust and gas in the solar vicinity
  and their star formation activities.
\end{abstract}


\section{Introduction}

Most molecular gas in our Galaxy is contained in molecular clouds, where the star formation takes places \citep{Blitz:1999}.
The formation of molecular clouds and the birth of stars are key processes in the life cycle of galaxies.
Understanding the physical properties of molecular clouds is thus of great importance.
Distance of the molecular clouds are fundamental to estimate from observations physical properties
including size and mass, etc. Moreover, mapping
the spatial distribution of molecular clouds at large scale plays
a crucial role for understanding their formation and evolution, and their role in star formation
\citep{Li2013, Li2016, Goodman2014, Wang2015}. However, building a large sample of molecular clouds
with accurate distance estimates is a difficult task.

The traditional method used to trace molecular clouds and derive their basic physical properties makes use of
CO observations. H$_2$, the dominant species in a molecular cloud, is hard to excite and observe under
typical conditions of dark, cold clouds. In comparison, heteronuclear diatomic molecule CO, the second most abundant
species in a molecular cloud, can be easily excited and observed, and has become the primary tracer of
molecular clouds. There are now many large-area CO surveys of the Milky Way and
a large number of molecular clouds have already been
identified (e.g. \citealt{Magani1985, Dame2001, May1997}). Distances of those clouds have been derived from the Galactic rotation curve
(e.g. \citealt{Brand1994, May1997, Nakagawa2005, Roman2009,  Garcia2014, Miville2017}).
However, those kinematic distances suffer from large uncertainties due to the difficulties of determining an accurate rotation
curve and the influence of the peculiar velocities and the non-circular motions,  as well as the well-know near-far 
ambiguity -- where one velocity can be related to two distances when a cloud is distributed in the inner Galaxy. 

A complementary approach to probe the properties of molecular clouds is to trace the molecular gas by dust
observations, specifically via the optical/near-infrared (IR) dust extinction measurements \citep{Goodman2009, Chen2017}.
Based on the two-dimensional (2D) Galactic extinction maps, \citet{Dobashi2011} have identified more than 7,000
molecular clouds. But they were unable to obtain distance informations for those molecular clouds given the 2D nature of the extinction maps.
\citet{Marshall2009} have obtained distances for over 1,000 clouds by analyzing the 3D extinction maps obtained by comparing
the observed colour distributions of Galactic giant stars with those predicted by the Galactic model. Their distance estimates have
uncertainties about 0.5-1\,kpc. \citet{Lada2009} and \citet{Lombardi2011} have obtained
distance estimates for a number of clouds by comparing the number of unextinguished stars, presumably
located in front of the clouds, with the predictions of the Galactic model. Recently,
due to the availability of large amounts of multi-band photometric and astrometric data, we are now able to obtain
values of distance and dust extinction for tens of millions individual stars
\citep{Chen2014, Chen2019, Green2015, Lallement2019}. Based on PanSTARRS-1 data,
\citet{Schlafly2014} present a catalogue of distances to molecular clouds selected from \citet{Magani1985} and
\citet{Dame2001} by the 3D extinction mapping method.
The distance estimates of \citet{Schlafly2014} have uncertainties of about 10\,per\,cent.
Using a similar technique, \citet{Zucker2019}
obtain distances to dozens of local molecular clouds using a combination of optical
and near-IR photometry and the Gaia Data Release 2 (Gaia DR2; \citealt{Gaia2018}) parallaxes \citep{Lindegren2018}.
Most recently, \citet{Yan2019} present a catalogue of distances
to molecular clouds at high Galactic latitudes based on estimates of parallax and extinction from Gaia DR2.
Benefiting from the large numbers of stars for the individual  molecular clouds and the robust estimates of the stellar
distances from Gaia DR2, the errors of distances obtained by \citet{Zucker2019} and
\citet{Yan2019} are typically only about 5\,per\,cent.

Most of the molecular clouds are located in the Galactic disk,
especially at Galactic low latitudes ($|b|$ $<$ 10\degr).
However, hitherto only a few local
well-studied molecular clouds at low latitudes, such as the California and Pipe Nebula,
have accurate distance estimates based on the 3D dust extinction mapping method. In the Galactic plane,
there are numerous high-density clouds and they tend to overlap with each others along the sightlines.
This poses extra difficulty to isolate the individual molecular clouds from their
foreground or background ones. In this work,
we identify and isolate the molecular clouds in the Galactic plane by applying a hierarchical cluster analysis technique
to the 3D colour excess maps of the Galactic disk presented by \citet[hereafter \ppi]{Chen2019}. For each
cloud, we select stars in the overlapping directions and fit the colour excess and distance relation for them.
The fit gives the distance to the cloud.

This paper is part of an ongoing project to study the interstellar dust and the Milky Way structure. The basic data are from \ppi\ that
presents 3D interstellar dust reddening maps of the Galactic plane (Galactic longitude 0\degr\ $<~l<$ 360\degr\ and latitude
$|b|$ $<$ 10\degr) in three colour excesses, \egk, \ebr \ and \ehk.
In this work, we will present a large catalogue  of 567 molecular clouds in the Galactic disk with  accurate determination of distances and other physical parameters.
With the data, we will carry out a detail analysis of the 3D motions of the molecular clouds and study the
kinematics of the Galactic spiral structure (Li et al. 2020, in preparation).

The paper is structured as following. In Section\,2, we describe the data. In Section\,3 we introduce our method for
isolating the molecular clouds in the disk and determining the distances to them. We present our results in Section\,4 and
discuss them in Section\,5. We summarize in Section\,6.

\section{Data}

In \ppi, we have calculated the values of colour excesses \egk, \ebr\ and \ehk \ of more than 56 million stars located at low Galactic
latitudes ($|b|$ $<$ 10\degr). In doing so, we combined the high quality optical photometry from Gaia DR2 \citep{Gaia2018} with the
near-IR photometry from the Two Micron All Sky Survey (2MASS; \citealt{Skrutskie2006}) and the Wide-Field Infrared Survey
Explorer (WISE; \citealt{Wright2010}).  The machine learning algorithm Random Forest Regression was applied to the
photometric data to obtain values of colour excesses of the individual stars, using a training data sample
constructed from spectroscopic surveys. The typical uncertainty of  the resulted colour excess values was about
0.07\,mag in $E(B-V)$ (\ppi). Distances estimated by \citet{Bailer2018}, who transfer the Gaia DR2 parallaxes to distances using
a simple Bayesian approach, were adopted for the stars. A simple cut of the Gaia DR2 parallax uncertainties of smaller than
20\,per\,cent was applied to exclude the stars with large distance errors. The analysis yielded a sample,
denoted  `C19 Sample', that consists of over 32 million stars
with estimates of distances and colour excesses. Based on the C19 Sample, \ppi\ presents 3D maps of colour excess \ebr\ of
the Galactic disk. The maps have an angular resolution of  6$^{\prime}$ and cover the entire Galactic disk. The typical depth limit
of the maps is about 4\,kpc  from the Sun.

\section{Method}\label{sec:method}
As the first step of our method, the 3D colour excess maps from \ppi\ are used to isolate the individual molecular clouds.
We adopt a cloud identification method that uses a clustering hierarchical algorithm
to identify coherent structures in the position ($l$, $b$) and distance ($d$) space at the same time.
We use the Python program Dendrogram \citep{Rosolowsky2008} to identify molecular clouds
in the 3D data cube of the \ebr\ maps.
   Dendrograms are tree representations of the hierarchical structure of nested
    isosurfaces in three-dimensional line data cubes. 
The algorithm has three input parameters:
i) ``min\_value'', the minimum value used to mask any structure that peaks below it; ii) ``min\_delta'', the
minimum significance for structures used to exclude any local maxima identified because
of the noise; and iii) ``min\_npix'', the minimum number of pixels that a structure should contain. In the current work,
we test with different parameters to optimise our selection. Finally we set, min\_value = 0.05\,mag\,kpc$^{-1}$,
min\_delta = 0.06\,mag\,kpc$^{-1}$ and
min\_npix = 20. The ``{\it leaves}'' of a Dendrogram correspond to the regions that have density enhancements, i.e. the
individual molecular clouds in our case.

For each molecular cloud ({\it leaf}), program Dendrogram provides its 3D spatial ranges ($l,~b$ and $d$).
We select all the C19 Sample stars that fall within the $l$ and $b$ ranges of the cloud.
The values of distance and \ebr\ colour excess
of the selected stars are then used to establish the colour excess and distance relation $E(d)$ along the sightline of this particular cloud.
Since the dust density in a molecular cloud is much higher than that in
the diffuse medium, one is
expected to find a sharp increasing of the colour excess values at the position of the cloud.
The second step of our method is to find the position, i.e. the distance of the molecular cloud.
In the current work, we assume that a cloud correspond 
to a region where the colour excess increases significantly.
The colour excess profile $E(d)$ within the distance range of the cloud can thus be described as,
\begin{equation}
  E(d)=E^{0}+E^{1}(d) {~~~~~~\rm if~} d_{\rm min} < d < d_{\rm max},
\end{equation}
where $E^{0}$ and $E^1(d)$ are respectively the colour excess contributed by the foreground dust 
and  the molecular cloud at distance $d$, while $d_{\rm min}$ and $d_{\rm max}$ are 
respectively the minimum and maximum distances of the molecular cloud given by program Dendrogram. 
As  molecular clouds span quite a large area in the sky, the colour excesses of the foreground/background stars are different. 
To take such effects into account, 
similar as in the work of \citet{Schlafly2014} and \citet{Zucker2019}, we start with the 3D colour excess map from \ppi. We adopt 
$E^0_n= f_1E_{n}(d_{\rm min})$, where $n$ is the index of stars, $f_1$ is  the scale factor, $E^0_n $ is the foreground colour excess,  and   
$E_{n}(d_{\rm min})$  is the colour excess at distance $d_{\rm min}$ from the Chen et al. map. 
Similar as in \citet{Chen2017} and \citet{Yu2019}, along the line of sight we assume 
a simple Gaussian distribution of dust in the cloud. The colour excess profile for the cloud $E^1(d)$ is then given by,
\begin{equation}
  E^{1}(d)=\frac{\delta E}{2}[1+ {\rm erf}\left(\frac{d-d_0}{\sqrt{2} \delta d}\right)],
\end{equation}
where $\delta E$ is the total colour excess contributed 
by the dust grains in the cloud, and $\delta d$ and $d_0$ are 
respectively the width (depth) and distance of the cloud. Again we adopt 
${\delta E_n} = f_2 [E_{n}(d_{\rm max})-E_{n}(d_{\rm min})]$, where  $\delta E_n$ is the total colour excess,  
$f_2$  is the scale factor,  $E_{n}(d_{\rm min})$ and $E_{n}(d_{\rm max})$ are respectively 
the colour excess at distances $d_{\rm min}$ and $d_{\rm max}$ from the Chen et al. map.
 
For each cloud,  the scale factors $f_1$ and $f_2$, and the width and distance of the cloud $\delta d$ and $d_0$ 
are free parameters to fit.  There could be more than one dust clouds for a given sightline. To avoid contamination by other clouds, 
we fit the colour excess profile only in a limited distance range, 
$d_{\rm min} - 0.2$\,kpc $<$ $d$ $<$ $d_{\rm max} + 0.2$\,kpc. 
The values of colour excess of the individual stars in the distance range are then fitted with the colour excess model 
described above with the IDL program MPFIT \citep{Markwardt2009}, which is
implementation of the Levenberg-Marquardt least-square optimization algorithm  \citep{More1978}. 
No priors are adopted for the fitting.
The uncertainties of the resultant parameters are derived using a Monte Carlo method.
For each molecular cloud, we randomly generate 300 samples of stars taking into account their
distance and colour excess uncertainties.
We apply the fitting algorithm and derive the parameters for all
samples. The results for each of the parameter follow a Gaussian distribution.
The dispersion of the Gaussian distribution is taken as the error of the
corresponding parameter.

Once the distances of the individual molecular clouds have been determined, in the third step of our method, we calculate their physical properties.
The solid angle subtended by an identified molecular cloud $\Omega$ is computed by integrating that of all pixels belonging to the cloud, i.e.,
\begin{equation}
  \Omega = \sum_{i=0}^{N} \Delta l  \Delta b  \cos b_i,
\end{equation}
where $i$ is the index of a pixel belonging the cloud, $\Delta l $ =  $\Delta b $ = 0.1\degr\ the angular width of the pixel and
$b_i$ the Galactic latitude of the $i$th pixel. Given the distance $d_0$ to the cloud, the area $S$ and the linear
radius $r$, respectively units of pc$^2$ and pc, of the cloud are given by,
\begin{equation}
  S = \Omega d_0^2,
\end{equation}
and
\begin{equation}
  r = \sqrt{\frac{S}{\pi}}.
\end{equation}
Finally, we estimate the mass
of each cloud, assuming that the dust is distributed as a thin sheet at its center position
\citep{Lombardi2011, Schlafly2015, Chen2017}, by,
\begin{equation}
  M = \dfrac{\mu m_{\rm H} d_0^2}{DGR}\sum_{i=0}^{N} {\delta A_{V,i}} \Delta l  \Delta b  \cos b_i,
\end{equation}
where $\mu$ is the mean molecular weight, $m_{\rm H}$ the mass of the hydrogen atom, $\delta A_{V,i}$ the  optical extinction of the cloud in 
the $i$th pixel and $DGR$ the dust-to-gas ratio,
\begin{equation}
  DGR = \dfrac{A_V}{N({\rm H})}=\dfrac{A_V}{N({\rm H~{\scriptstyle I}})+2N({\rm H_2})}.
\end{equation}
In the current work, we adopt $\mu$ = 1.37 \citep{Lombardi2011} and
$DGR$ = 4.15 $\times$ 10$^{-22}$\,mag\,cm$^2$ \citep{Chen2015}.
The V-band extinction $A_{V,i}$ in pixel $i$ of a given molecular cloud is converted from its colour excess $\delta E_i$ [see Eq.~(2)]
using the extinction law of \ppi,
assuming $R_V$ = 3.1. This yields,
\begin{equation}
  \delta A_{V,i} = 2.33\delta E_i(G_{\rm BP}-G_{\rm RP}).
\end{equation}
The surface mass density $\Sigma$ of the molecular cloud, in units of $M_{\odot}$\,pc$^{-2}$, is then calculated as
\footnote{Where the surface mass density is linearly proportional to $A_V$.} ,
\begin{equation}
  \Sigma = \dfrac{M}{S}.
\end{equation}

\section{Results}

\begin{figure*}
  \centering
  \includegraphics[width=0.99\textwidth]{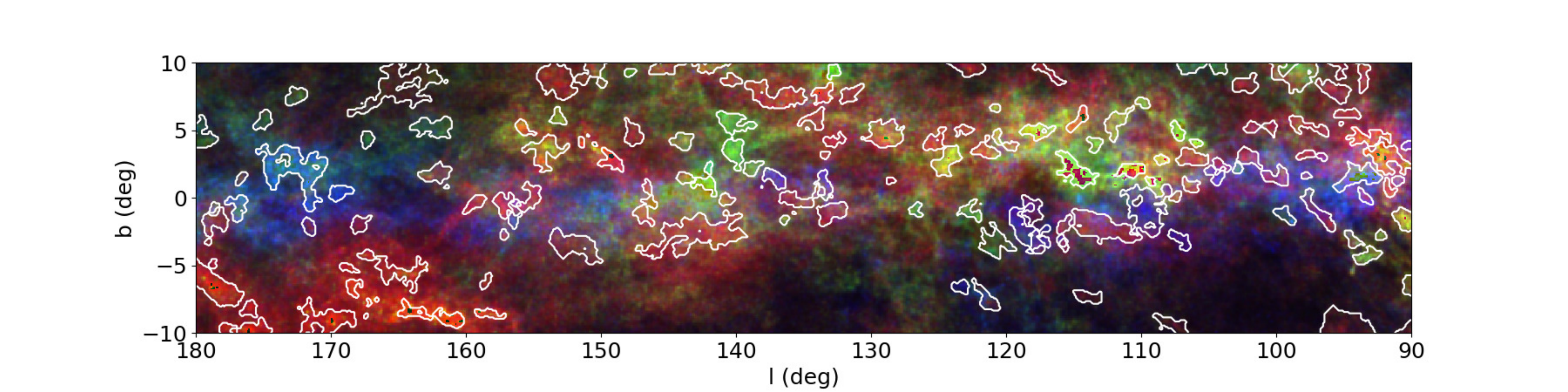}
  \includegraphics[width=0.99\textwidth]{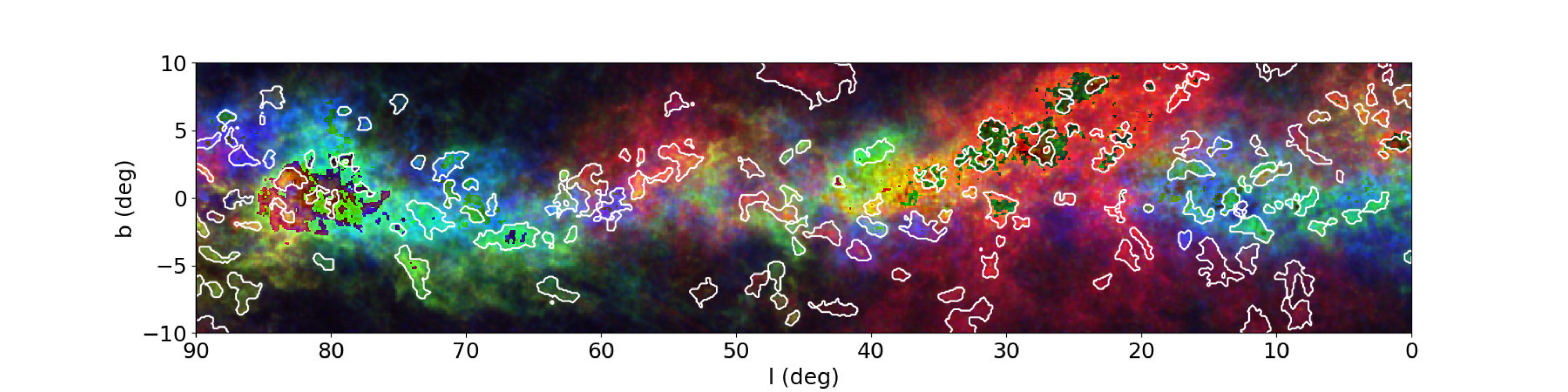}
  \includegraphics[width=0.99\textwidth]{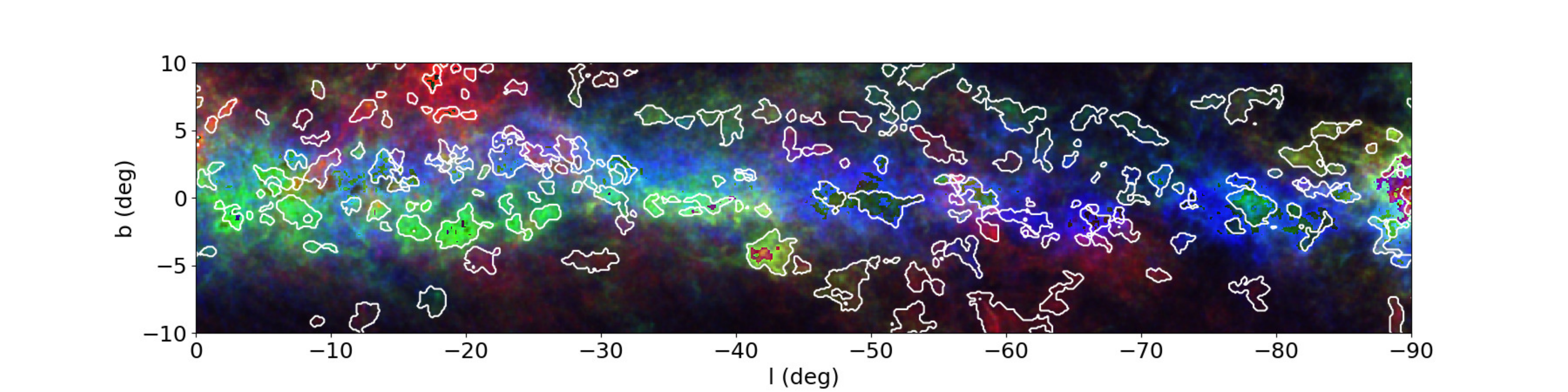}
  \includegraphics[width=0.99\textwidth]{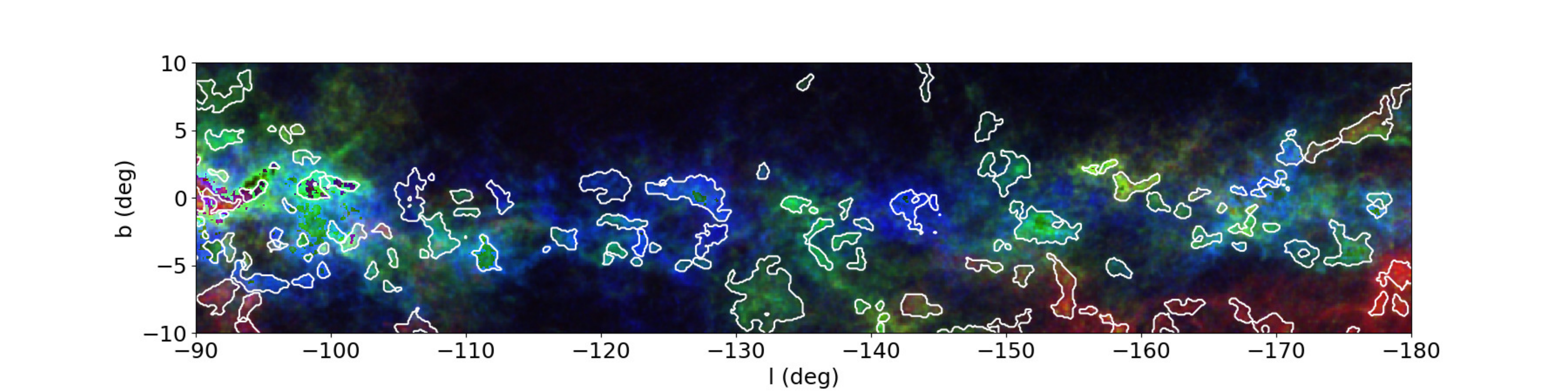}
  \caption{Distributions of the molecular clouds identified in the current work in the Galactic coordinates. The background
    three-colour composite images illustrate the three-dimensional distributions of dust in the Galactic disk from \ppi. Red, green and blue scales
    show the distributions of the colour excess $\delta$\ebr\ in distance slices 0 - 1000\,pc, 1000 - 2000\,pc and 2000 - 5000\,pc from the
    Sun, respectively. The white polygons mark the $l$ and $b$ boundaries of the individual molecular clouds identified in the current work.}
  \label{dend}
\end{figure*}

\begin{figure*}
  \centering
  \includegraphics[width=0.99\textwidth]{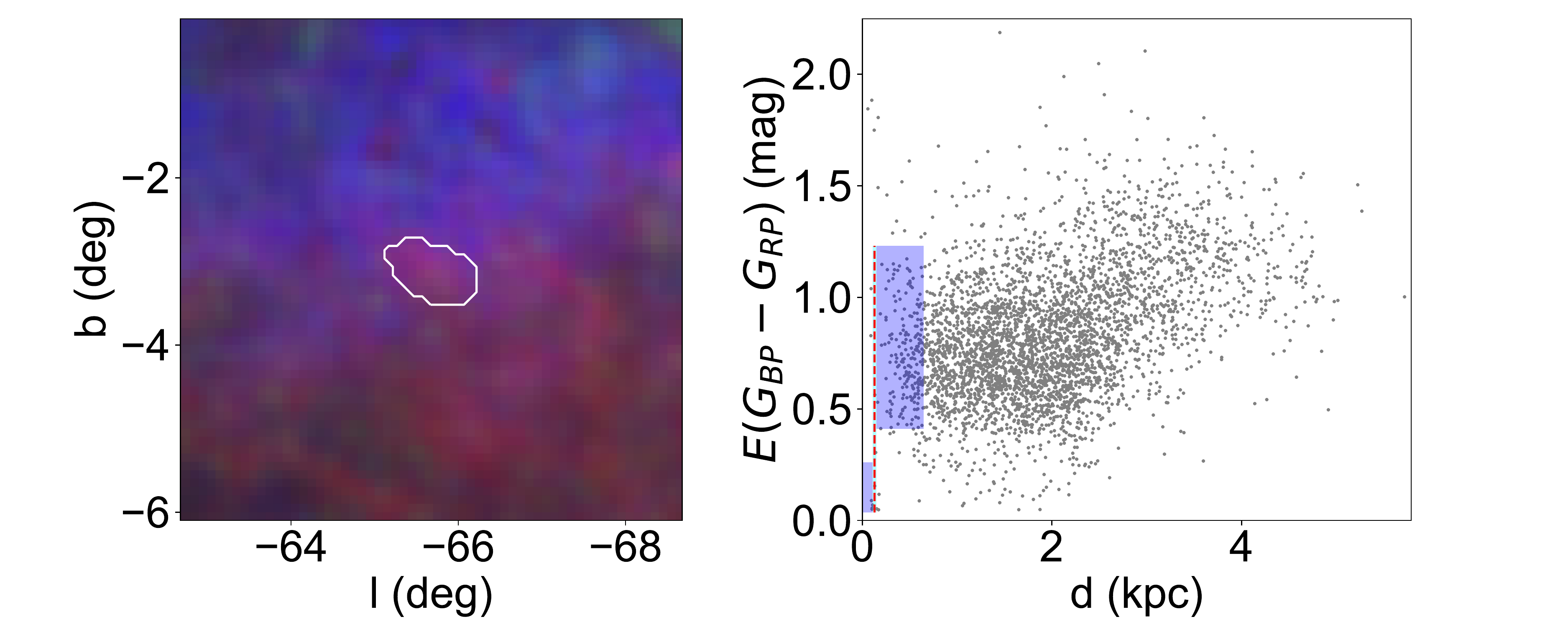}
  \includegraphics[width=0.99\textwidth]{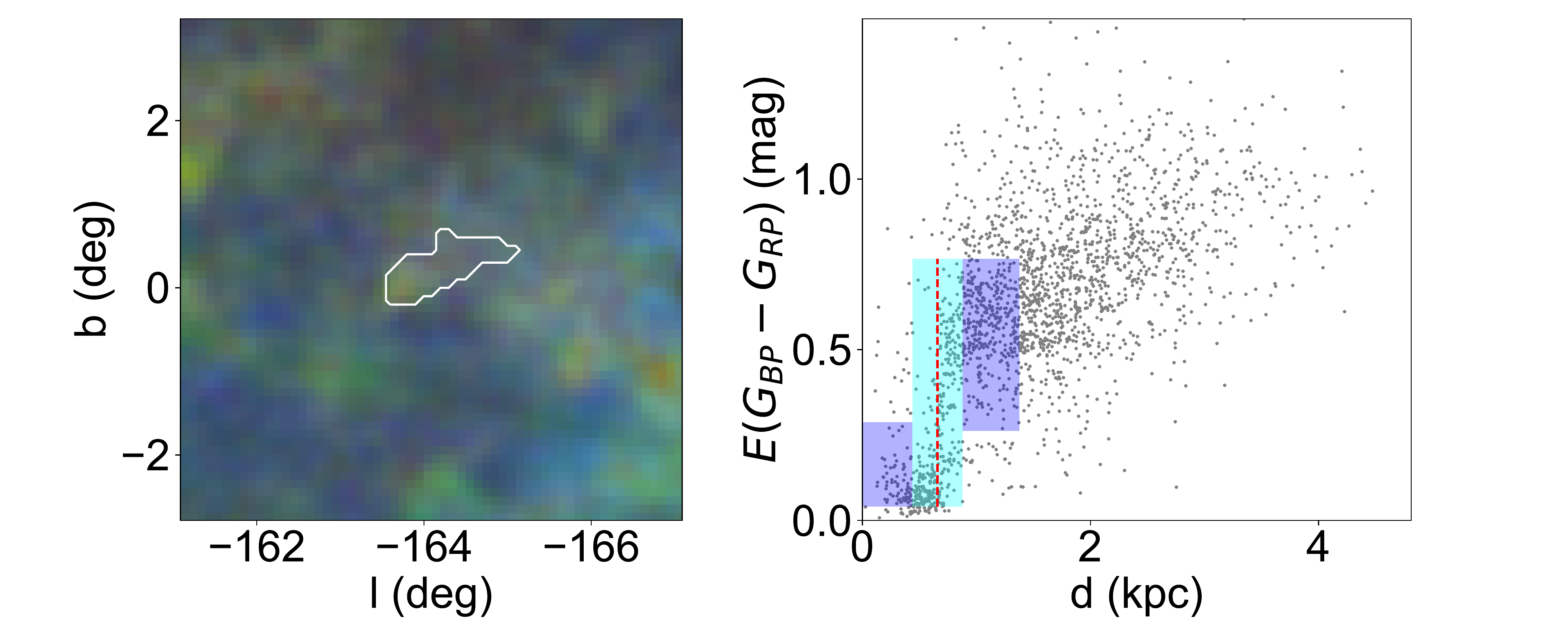}
  \includegraphics[width=0.99\textwidth]{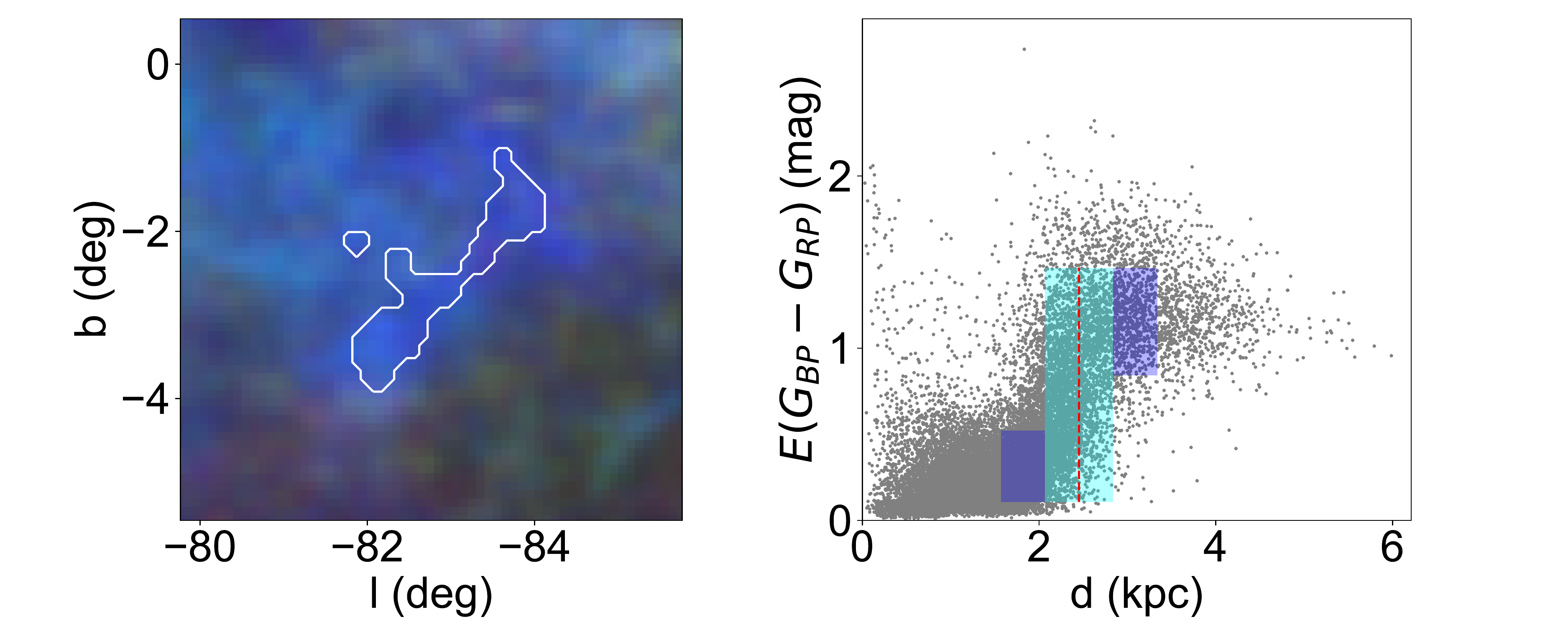}
  \caption{Distances to clouds No.~61 (upper panels), 318 (middle panels) and 539 (bottom panels).
    In the left panels, red, green and blue scales
    show the distributions of the colour excess $\delta$\ebr\ in distance slices 0 - 1000\,pc, 1000 - 2000\,pc and 2000 - 5000\,pc from the
    Sun, respectively.
    White polygons show the region of sky of the
    clouds and from which the stars used in the distance determination are drawn.
    In the right panels, the values of colour excess \ebr\ of the selected stars
    are plotted against distances of the stars. The rectangles show the best-fit colour excess profiles.
   In each panel, the cyan rectangle represents the width of the cloud ($\delta d$) and the blue rectangles
    show the 1$\sigma$ regions  of the foreground and background colour excesses of the cloud.   
    The red dashed lines show the distances of the clouds.}
  \label{disget}
\end{figure*}

\begin{figure*}
  \centering
  \includegraphics[width=0.99\textwidth]{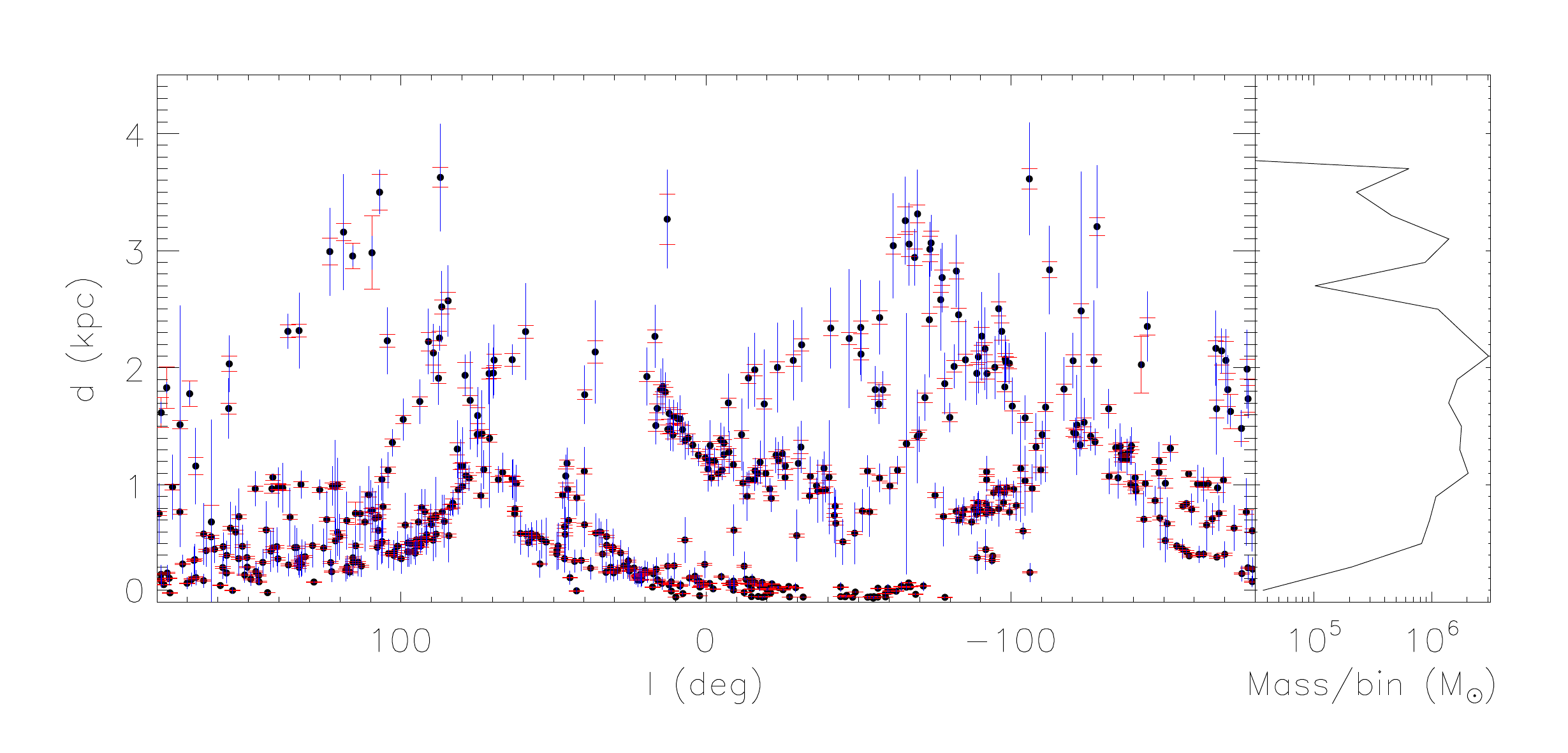}
  \caption{Derived cloud distances plotted against their Galactic longitudes.
    The red and blue error-bars show the distance uncertainties and the derived width/depth ($\delta d$) of the clouds, respectively.
    Also shown to the right of the panel is the distribution of the estimated total masses of the molecular clouds in the individual distance bins
    of width 0.2\,kpc. }
  \label{ldis}
\end{figure*}

\begin{table*}
  \centering
  \caption{Catalogue of molecular clouds.}
  \begin{tabular}{rrrr  *{6}{r} }
    \hline
    \hline
    ID  & $l$        & $b$       & $\Omega$  & $r$  & $d_0$            & $\delta d$      & $\overline{E(G_{\rm BP}-G_{\rm RP})}$  & $M$           & $\Sigma$               \\
        & ($\degr$)  & ($\degr$) & (deg$^2$) & (pc) & (pc)             & (pc)            & (mag)          & ($M_{\odot}$) & $ (M_{\odot}$\,pc$^2$) \\
    \hline
  1  &  177.727  &   $-$9.596  &    0.730  &          1.3   &        149.3$\pm$3.5   &         10.2$\pm$0.2   &     0.80   &      248.5   &   50.1   \\
  2  &  175.715  &   $-$9.651  &    1.903  &          1.1   &         78.9$\pm$1.9   &         10.0$\pm$0.2   &     0.97   &      218.5   &   60.6   \\
  3  &  170.131  &   $-$8.972  &    5.749  &          3.8   &        160.8$\pm$3.8   &         51.6$\pm$1.2   &     0.99   &     2787.5   &   61.5   \\
  4  &  107.778  &   $-$9.278  &    2.517  &          7.3   &        467.8$\pm$11.0   &        192.1$\pm$4.5   &     0.36   &     3723.7   &   22.2   \\
  5  &   94.961  &   $-$9.339  &    5.607  &         12.4   &        531.7$\pm$12.5   &        128.2$\pm$5.5   &     0.37   &    11205.9   &   23.2   \\
  6  &   32.797  &   $-$8.898  &    4.773  &          5.5   &        254.3$\pm$6.0   &         90.6$\pm$2.1   &     0.64   &     3746.5   &   39.9   \\
  7  &   11.421  &   $-$9.357  &    2.704  &          2.8   &        173.1$\pm$4.1   &         52.2$\pm$1.2   &     0.51   &      787.4   &   31.9   \\
  8  &  $-$52.953  &   $-$9.019  &    4.839  &          0.9   &         41.7$\pm$1.0   &         10.0$\pm$0.2   &     0.33   &       52.9   &   20.6   \\
  9  &  $-$59.581  &   $-$8.908  &    8.982  &          3.5   &        117.5$\pm$2.8   &         38.3$\pm$0.9   &     0.47   &     1103.2   &   29.2   \\
 10  &  $-$93.764  &   $-$9.628  &    1.242  &          3.9   &        355.6$\pm$8.4   &         40.7$\pm$4.0   &     0.52   &     1540.3   &   32.2   \\
    ... & ...        & ...       & ...       & ...  & ...              & ...             & ...            & ...           & ...                    \\
560  &  $-$68.358  &   $-$1.273  &    0.760  &         25.3   &       2942.1$\pm$69.4   &        241.1$\pm$83.1   &     0.60   &    75291.7   &   37.6   \\
561  &   12.740  &   $-$1.095  &    0.400  &         20.4   &       3269.0$\pm$214.5   &        422.1$\pm$142.0   &     0.36   &    28886.7   &   22.2   \\
562  &  $-$73.360  &   $-$0.551  &    0.420  &         19.2   &       3012.1$\pm$109.1   &        233.4$\pm$111.1   &     0.55   &    39587.5   &   34.1   \\
563  &  $-$73.759  &    0.128  &    0.350  &         17.9   &       3065.8$\pm$98.7   &        239.2$\pm$108.8   &     0.56   &    34729.9   &   34.7   \\
564  &  $-$69.328  &    1.674  &    1.139  &         34.8   &       3313.7$\pm$78.2   &        377.1$\pm$61.7   &     0.27   &    64708.4   &   17.0   \\
565  & $-$105.961  &    0.176  &    5.859  &         86.1   &       3612.7$\pm$85.3   &        481.5$\pm$36.6   &     0.33   &   473619.1   &   20.3   \\
566  &  $-$65.300  &   $-$0.325  &    0.410  &         20.5   &       3256.2$\pm$117.4   &        374.5$\pm$79.6   &     0.56   &    46280.5   &   35.0   \\
567  &   87.082  &  $-$0.151  &    1.830  &         48.3   &       3625.5$\pm$85.6   &        460.9$\pm$38.8   &     0.36   &   166119.2   &   22.7   \\
    \hline
  \end{tabular}
  \parbox{\textwidth}{\footnotesize \baselineskip 3.8mm
    The Table is available in its entirety in machine-readable form in the online version of this manuscript and also at the website \\
    ``http://paperdata.china-vo.org/diskec/dustcloud/table1.txt''.}
\end{table*}

In the current work, we have identified 567 molecular clouds with program Dendrogram.
The distribution of the clouds, catalogued in Table~1, in Galactic coordinates $l$ and $b$ is presented in Fig.~\ref{dend}.
The typical solid angle subtended by a cloud is about 1.5\,deg$^2$.
Typically there are $\sim$4,500 stars along the sightlines toward a cloud in the C19 Sample.

\subsection{Distances of the molecular clouds}

Fig.~\ref{disget}  shows the results of our colour excess profile fitting procedure for three example molecular clouds.
In general, the values of colour excess \ebr\  of stars in individual distance bins show significant dispersions, 
as those stars cover quite a large area ($\sim$ 1 $-$ 2\,deg$^2$). However, the “jumps” in colour excess values produced 
by the molecular clouds in the corresponding distance ranges given by program Dendrogram are clearly visible. 
They are nicely fitted by our colour excess profiles.

The panels in the first row of Fig.~\ref{disget} show the fitting analysis of a nearby molecular
cloud with distance $d$ of 127\,pc from the Sun. Due to the
saturation of the photometric data used in \ppi, there are only a few stars in the C19 Sample that are located in front of the cloud.
The quality of colour excess profile fit for the cloud is relatively poor compared to that of more distant clouds.

The panels in the second and third rows
of Fig.~\ref{disget} show examples of two clouds located respectively at distances
659\,pc and 2,453\,pc from the Sun.
Compared to the nearby molecular cloud, the distance ranges of the colour excess jumps of more distant
molecular clouds are larger.
As a result, more distant clouds have larger `widths' ($\delta d$).
This is mainly caused by the larger distance errors of stars at further distances.
If one assumes that the molecular
clouds are spherical, the intrinsic widths of the clouds range between $\sim$ 1-100\,pc (see Sect.~4.2).
On the other hand, the typical uncertainties of the Gaia parallaxes
for stars at distances of 2\,kpc are $\sim$10\,per\,cent, i.e. $\sim$ 200\,pc, which is twice of the
upper limit of the intrinsic widths of the clouds. Therefore, the resulted values of width $\delta d$ for distant clouds are mainly contributed by the
distance errors of the individual stars. However, benefiting from the large numbers of stars for the individual clouds,
the large distance uncertainties do not
have significant impacts on the distance determinations of the clouds.

In directions toward some molecular clouds, we are able to see more than one colour excess jump. The additional jumps are produced  by the
foreground or background clouds. With the distance ranges provided by
Dendrogram, we are able to exclude the contamination of those  foreground or background clouds and obtain accurate distances for the individual
identified clouds.

For each molecular cloud identified in the current work, we have made figures
analogous to Fig.~\ref{disget}. They are available online\footnote{http://paperdata.china-vo.org/diskec/dustcloud/allcloud.pdf}.
The best-fit values and uncertainties of distances $d_0$ and width $\delta d$, and the 
averaged colour excess $\overline{E(G_{\rm BP}-G_{\rm RP})}$ of the individual catalogued clouds
are listed in Table~1. Fig.~\ref{ldis} plots the centre longitude $l$ of all the molecular clouds
against their distances $d_0$.
The associated statistical distance uncertainties and the estimated widths (depths) of the clouds ($\delta d$) are also overplotted.
The cloud range in distances from $d_0\approx30$\,pc to $\sim 4,000$\,pc.
The catalogue should be complete between 1 and 3\,kpc. For distances larger than  3\,kpc, our sample is
incomplete due to the limited depth of the adopted 3D colour excess maps, while for distances closer than 1\,kpc, it is incomplete mainly
due to the incomplete coverage of Galactic latitudes of the dataset.
In addition, we note that the far-away clouds would appear be stretched due to the relatively large distance errors. Their observed differential 
colour excesses [$d$\ebr$/ds$, where $s$ is the distance] are actually smaller than the intrinsic values. 
Thus the far-away clouds with low dust density could be under our detection 
threshold. At larger distances, our selection procedure might have missed those low density clouds or the low density parts of the clouds. 
Owing to the accurate distances of stars yielded by the Gaia parallaxes and the large numbers  of stars available to trace the individual clouds, we
have achieved a very high precision for our distance estimates of the clouds. Most of the clouds in our catalogue have relative distance
uncertainties smaller than 5\,per\,cent. The typical `width (depth)' of the catalogued clouds is about 18\,per\,cent of
the distance values, largely caused by the distance errors of the individual stars  used to trace the clouds.

\subsection{Physical properties}

\begin{figure}
  \centering
  \includegraphics[width=0.48\textwidth]{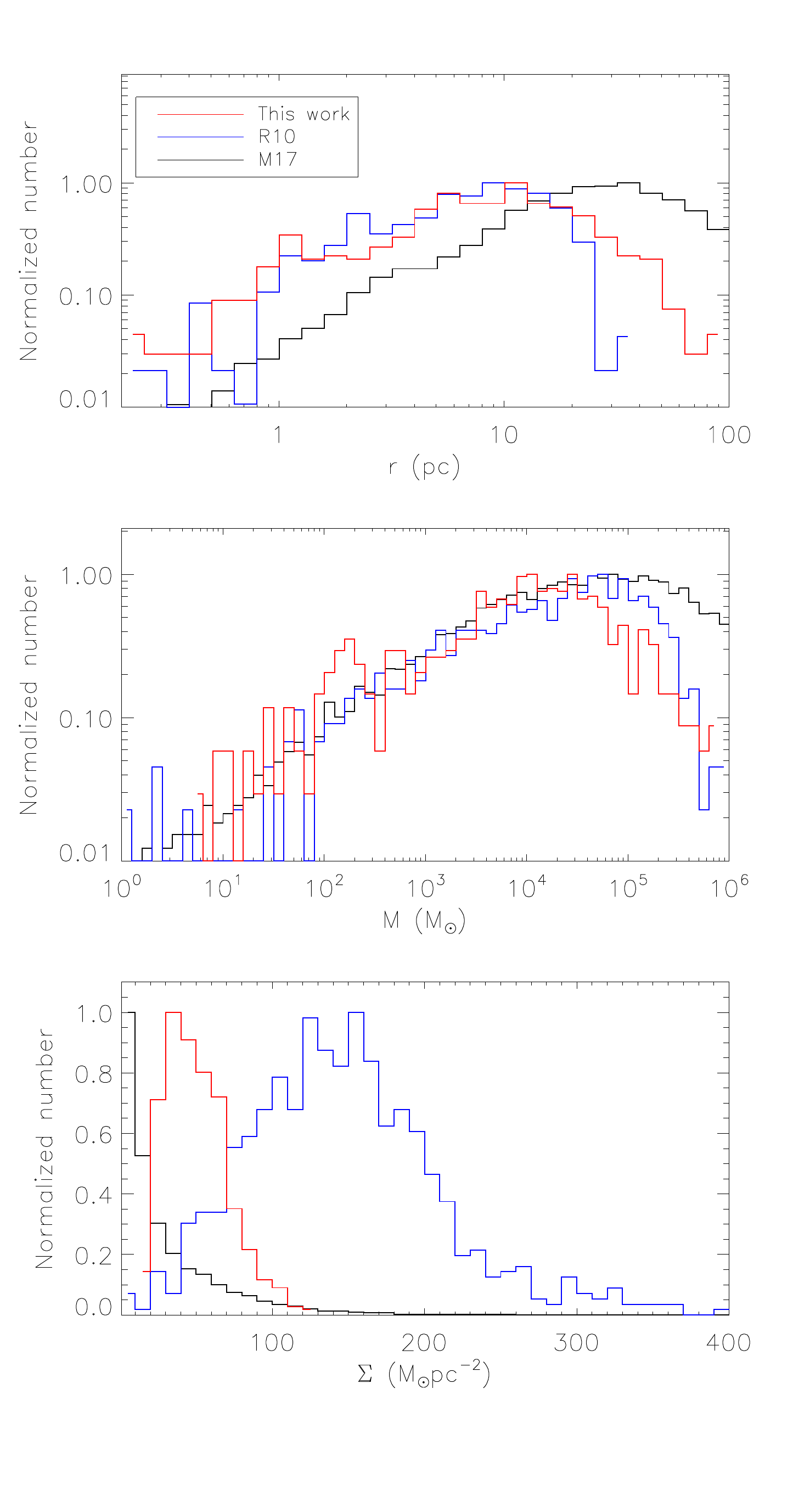}
  \caption{Histograms of the physical properties radii (upper panel),  masses (middle panel) and
    surface mass densities (bottom panel) of the molecular clouds catalogued in the current work (red lines), and those in
    \citet[blue lines]{Roman2010} and \citet[black lines]{Miville2017}.}
  \label{histos}
\end{figure}

\begin{figure}
  \centering
  \includegraphics[width=0.48\textwidth]{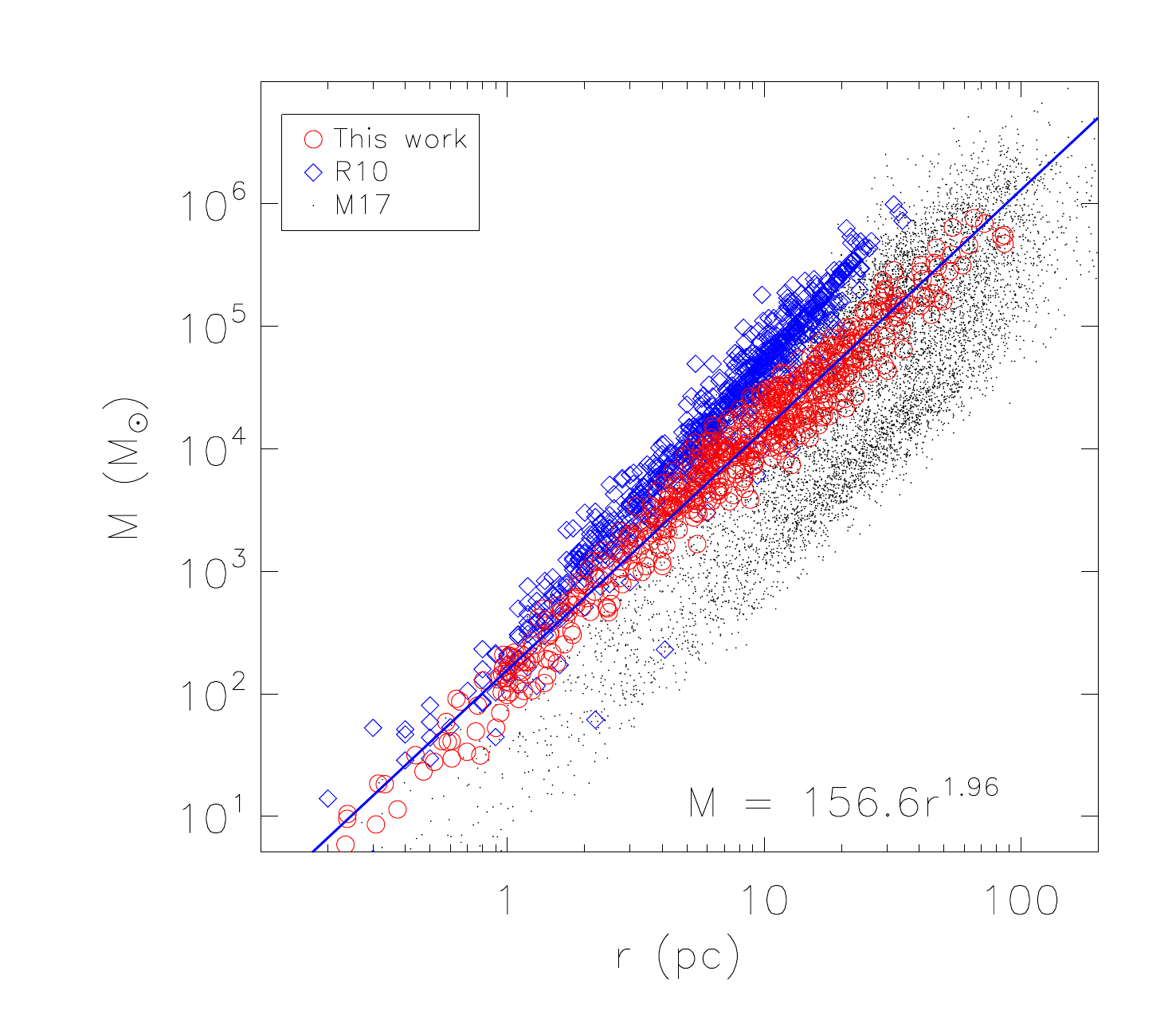}
  \caption{The masses plot against the radii for the molecular clouds in the current work (red circles),
    that of \citet[blue squares]{Roman2010} and \citet[black dots]{Miville2017}. The blue line is the best-fit correlation of the mass
    of the molecular clouds in our work and their radii. }
  \label{rmass}
\end{figure}

The derived physical properties, such as the physical radius $r$, mass $M$ and surface mass density $\Sigma$, of
the molecular clouds are also listed in Table.~1. Fig.~\ref{histos} plots the histograms of those
properties. The radii of the clouds range between
0.2 and 90\,pc with a median value of $\sim$ 8\,pc,
the masses between 5 and 700,000\,$M_{\odot}$ with a median value of
$\sim$ 10,000\,$M_{\odot}$ and the surface mass densities between
10 and 125\,$M_{\odot}$\,pc$^2$ with a median value of $\sim$ 50\,$M_{\odot}$\,pc$^2$.

The distributions of the corresponding properties of clouds
studied by \citet{Roman2010} and \citet{Miville2017} are also overplotted in Fig.~\ref{histos} for comparison.
\citet{Roman2010} have derived physical properties of 580 molecular clouds based on the
CO observations of the University of Massachusetts-Stony Brook (UMSB) and Galactic Ring surveys.
\citet{Miville2017} have isolated 8,107 molecular clouds from the CO observations of \citet{Dame2001} and derived their
physical properties. The molecular clouds of \citet{Roman2010} and \citet{Miville2017} cover
disk regions which are of much further distances from the Sun than those catalogued here. In addition, their
physical properties are obtained from data with methods much different from ours.
Nevertheless, the resultant distributions of the physical radii and masses of the molecular clouds are very similar,
suggesting that the structures identified in the current work are
essentially the same type of object as those identified in those previous studies.

Fig.~\ref{rmass} shows the correlation between the radii and masses of the molecular clouds.
The radii and masses of our molecular clouds are tightly correlated by a power-law, $M = 156.6 R^{1.96}$.
The exponent index 1.96, which corresponds to a typical surface density of $\sim$ 50\,$M_{\odot}$\,pc$^2$,
is slightly smaller than that found by \citet{Roman2010} but consistent with that of \citet{Miville2017}.
{This limiting surface density is largely caused by the fact that we have imposed a limiting extinction to our sample sources (Sect. \ref{sec:method}).}

The estimation of cloud masses and radii depends on how the cloud boundaries are defined.
The differences in cloud mass-size distributions reported in different work are most likely
caused by the different criteria used to define the clouds.

\begin{figure*}
  \centering
  \includegraphics[width=0.8\textwidth]{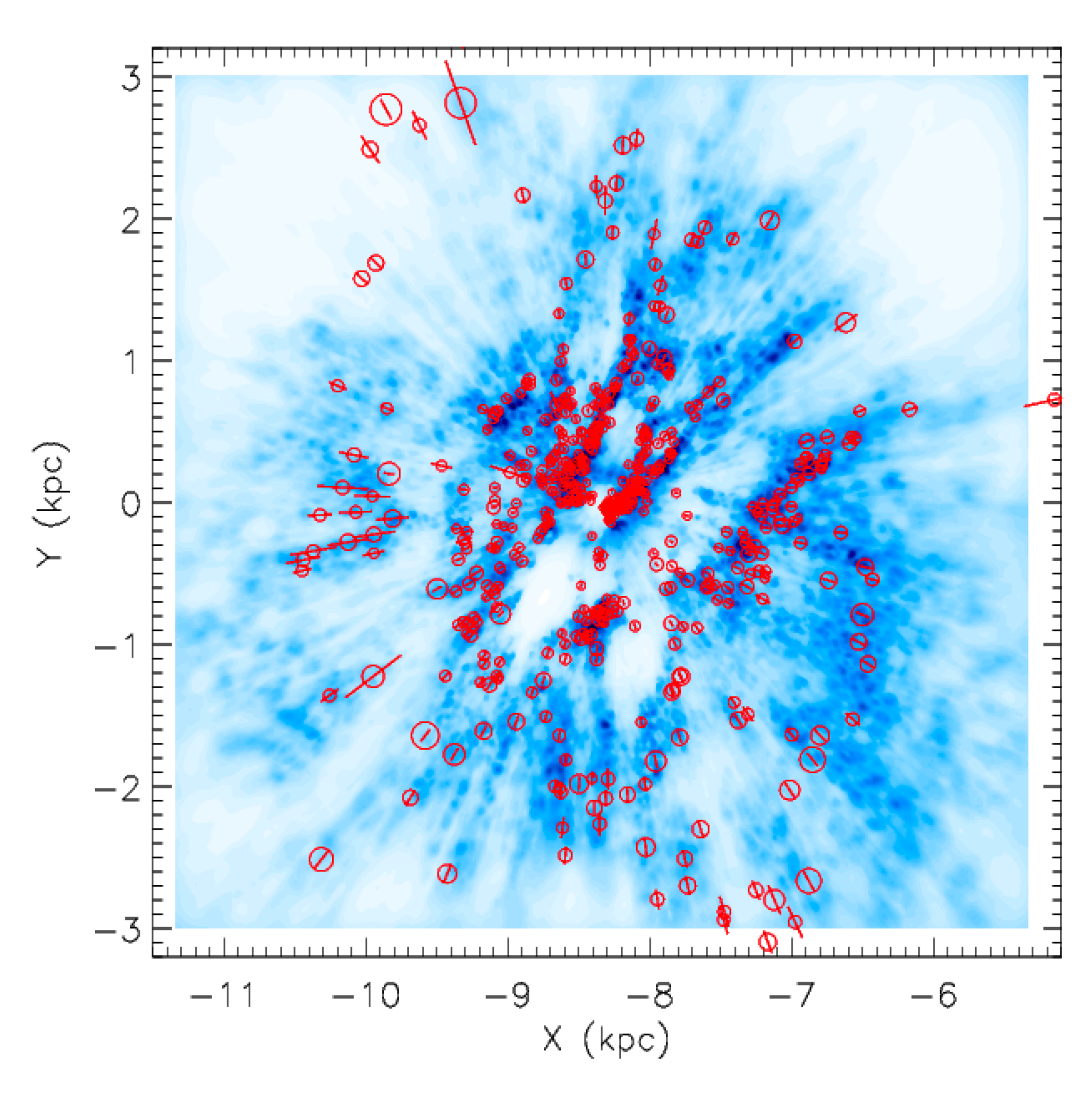}
  \caption{Spatial distribution of the molecular clouds identified and catalogued in the current work (circles)
    in the $X$-$Y$ plane, plotted over the distribution
    of the Galactic dust of disk vertical height $|Z|$ $<$ 0.4\,kpc (blue
    scales; \citealt{Lallement2019}). The sizes of the circles are proportional to the physic sizes of
    the individual clouds and the error-bars indicate their
    distance uncertainties.
    The Sun, assumed to be at the position of ($X,~Y,~Z$) = ($-$8.34, 0, 0)\,kpc, is located at the centre of the plot.}
  \label{l19}
\end{figure*}

\begin{figure*}
  \centering
  \includegraphics[width=0.75\textwidth]{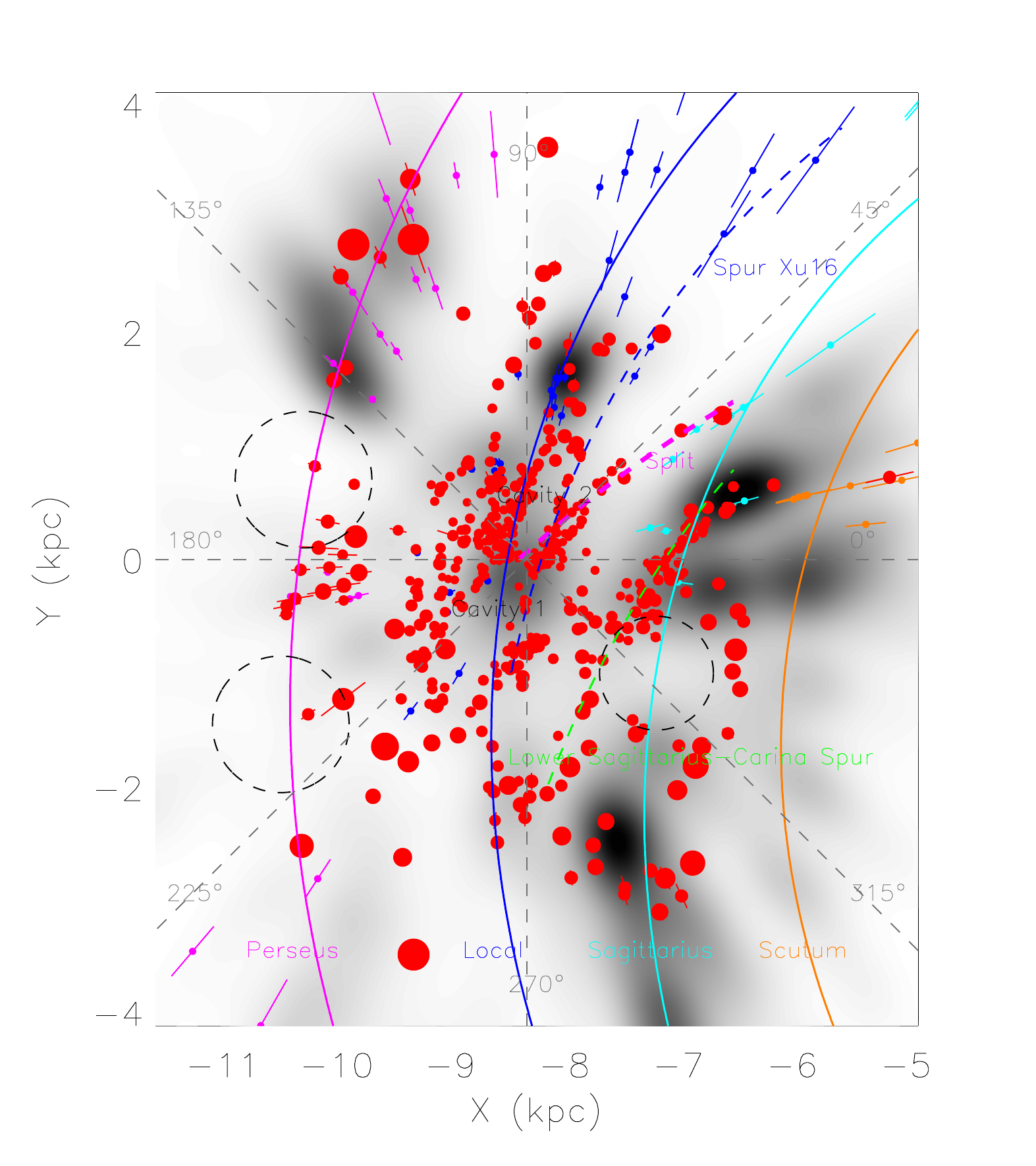}
  \caption{Spatial distribution of the identified molecular clouds in the $X$-$Y$ plane.
    The sizes of the circles are proportional to the physic sizes of
    the individual clouds and the error-bars indicate their
    distance uncertainties. Solid orange, cyan, blue and pink
    lines delineate the best-fit spiral arm models of the Sagittarius, Local and Perseus Arms presented in
    \citet{Chen2019b}.  Circles of the aforementioned colours are the masers from \citet{Xu2018b} that are probably
    associated with the individual Arms.
    The grey background image represents the spatial distribution of the OB stars from \citet{Chen2019b}.
    The green, pink and blue dashed lines mark respectively a possible spur `Lower Sagittarius-Carina Spur', 
    the `split' found by \citet{Lallement2019} and the spur identified by \citet{Xu2016}.
    Three black dashed circles mark the positions of three possible ‘hole’
    patterns in the Sagittarius and Perseus Arms identified in \citet{Chen2019b}.
    The Sun, assumed to be at the position of ($X,~Y,~Z$) = ($-$8.34, 0, 0)\,kpc, is located at the centre of the plot. The directions of
    $l$ = 0\degr, 45\degr, 90\degr, 135\degr, 180\degr, 225\degr, 270\degr\ and 315\degr\ are also marked in the plot.}
  \label{xydist}
\end{figure*}

\section{Discussion}

\begin{table*}
  \centering
  \caption{Comparison of distances of molecular clouds with the literature values.}
  \begin{tabular}{rrrrrr} 
    \hline
    \hline
    Name        & ID  & $l$       & $b$       & $d_0$            & Literature $d_0$                                     \\
                &     & ($\degr$) & ($\degr$) & (pc)             & (pc)                                                 \\
    \hline
    California  & 187 & 161.101   & -8.922    & 451.2$\pm$10.6    & 410$\pm$41$^a$, 470$\pm$26$^b$,450$\pm$23$^c$        \\
    California  & 189 & 158.282   & -8.578    & 473.2$\pm$11.2    & 410$\pm$41$^a$, 470$\pm$26$^b$,450$\pm$23$^c$        \\
    California  & 235 & 163.883   & -8.380    & 439.3$\pm$10.4    & 410$\pm$41$^a$, 470$\pm$26$^b$,450$\pm$23$^c$        \\
    Cepheus     & 353 & 106.260   & 9.687     & 1047.8$\pm$24.7  & 900$\pm$90$^a$, 923$\pm$47$^b$, 1043$^{+6}_{-7}$$^d$ \\
    Hercules    & 153 & 44.593    & 8.882     & 208.7$\pm$4.9    & 200$\pm$30$^a$, 227$\pm$12$^b$, 229$^{+4}_{-3}$$^d$  \\
    Pipe Nebula & 115 & 0.891     & 4.131     & 155.2$\pm$3.7    & 130$\pm$15$^e$, 145$\pm$16$^f$                       \\
    Pipe Nebula & 128 & -1.200    & 5.446     & 159.5$\pm$3.8   & 130$\pm$15$^e$, 145$\pm$16$^f$                       \\
    Pipe Nebula & 143 & -2.324    & 6.551     & 113.3$\pm$2.7   & 130$\pm$15$^e$, 145$\pm$16$^f$                       \\
    Circinus    & 242 & -42.343   & -4.054    & 819.2$\pm$19.3    & 700$\pm$350$^g$                                      \\
    CMa OB1     & 366 & -137.931  & -3.109    & 1217.8$\pm$28.7   & 1456$\pm$146$^a$, 1209$\pm$64$^b$,1150$\pm$64$^h$    \\
    CMa OB1     & 402 & -136.195  & -4.325    & 1225.8$\pm$28.9  & 1456$\pm$146$^a$, 1209$\pm$64$^b$,1150$\pm$64$^h$    \\
    CMa OB1     & 414 & -136.119  & -2.153    & 1265.3$\pm$29.9   & 1456$\pm$146$^a$, 1209$\pm$64$^b$,1150$\pm$64$^h$    \\
    Maddalena   & 529 & -142.675  & -1.063    & 2026.4$\pm$243.4 & 2350$\pm$235$^a$, 2072$\pm$110$^b$                   \\
    Maddalena   & 544 & -144.632  & -0.413    & 2353.4$\pm$73.7  & 2350$\pm$235$^a$, 2072$\pm$110$^b$                   \\
    Mon R2      & 233 & -147.177  & -9.806    & 866.7$\pm$20.5    & 952$\pm$95$^a$, 778$\pm$42$^b$, 905$\pm$37$^h$       \\
    Mon OB1     & 319 & -159.166  & 0.885     & 791.0$\pm$18.7    & 890$\pm$89$^a$, 745$\pm$40$^b$                       \\
    Orion Lam   & 12  & -170.023  & -9.455    & 408.4$\pm$9.6    & 427$\pm$43$^a$, 402$\pm$21$^b$, 445$\pm$50$^h$       \\
    Orion Lam   & 14  & -161.521  & -9.236    & 406.9$\pm$9.6   & 427$\pm$43$^a$, 402$\pm$21$^b$, 445$\pm$50$^h$       \\
    Orion Lam   & 25  & -163.075  & -8.053    & 409.2$\pm$9.7   & 427$\pm$43$^a$, 402$\pm$21$^b$, 445$\pm$50$^h$       \\
    Orion Lam   & 188 & -167.414  & -8.462    & 383.9$\pm$9.1    & 427$\pm$43$^a$, 402$\pm$21$^b$, 445$\pm$50$^h$       \\
    \hline
  \end{tabular}
  \parbox{\textwidth}{\footnotesize \baselineskip 3.8mm
    References: $^a$\citet{Schlafly2014}, $^b$\citet{Zucker2019},
    $^c$ \citet{Lada2009}, $^d$\citet{Yan2019}, $^e$\citet{Lombardi2006}, $^f$\citet{Alves2007}, $^g$\citet{Bally1999},
    $^h$\citet{Lombardi2011}.
  }
\end{table*}

\subsection{Comparison with previous work}

The unique catalogue presented in the current work is one of the largest homogeneous
catalogues of molecular clouds with accurate distance estimates.
The clouds  span in
distances from $\sim$30 to $\sim$4,000\,pc, with typical uncertainties of less
than 5\,pre\,cent. To verify the robustness of our distance estimates,  we have selected some well studied giant molecular
clouds and collect their distance estimates from the literature.
The results are listed in Table~2.
Overall, our values agree well with those from the literature.

For some nearby giant molecular cloud complexes, such as the California,
Pipe Nebula, CMA OB1 and Orion Lam, we have isolated
more than one clouds in each of them. The resulted distance estimates
of the individual clouds are consistent with each others within 3$\sigma$.
There are also some molecular clouds that have extended to the medium or high Galactic latitudes, such as the
Orion Lam, Mon R2, Hercules, etc. We have been only able to identify parts
of them, limited by the latitude coverage of our data ($b$ $<$ 10\degr). Nevertheless, our
estimated  distances are in good agreement with those in the literature.
The Cepheus Flare, for example, is a structure complex that contains at least two components \citep{Kun2008, Schlafly2014}.
As the footprint of our data are limited to Galactic latitudes smaller than 10\degr,
we are only able to identified parts of its southern component, named cloud No.~353 in the current work.
It is centered at $l$ = 106.260\degr\ and $b$ =  9.687\degr\ and
have a distance of 1047.8$\pm$24.7\,pc. Literature
estimates of distance for this component are 900$\pm$90\,pc \citep{Schlafly2014}, 923$\pm$47\,pc \citep{Zucker2019} and 1043$^{+6}_{-7}$\,pc \citep{Yan2019},
all in good agreement with our current result.
The dispersion of differences between our distance estimates and those from 
\citet{Zucker2019} and \citet{Yan2019} is about 5.6\,per\,cent.  
Note that all the cloud distances here are derived from Gaia DR2 parallaxes.

We also compare the spatial distribution of our molecular clouds to the 3D dust distribution of the
Galactic plane from \citet{Lallement2019}. Based on the photometric data of Gaia DR2 and 2MASS, and
the parallaxes of Gaia DR2, \citet{Lallement2019} present the 3D distribution of the interstellar dust in a volume of
6 $\times$ $6$ $\times$ 0.8\,kpc$^3$ around the Sun with
a hierarchical inversion algorithm. The comparison is shown in Fig.~\ref{l19}.
The spatial distribution of the molecular clouds catalogued here
is in excellent agreement with the dust distribution of \citet{Lallement2019}. The good agreement
validates the robustness of methods used in both papers.

\subsection{The Galactic spiral structure as traced by the molecular clouds}

As Figs.~\ref{ldis} and \ref{l19} show, the molecular clouds in our catalogue
trace clearly the large-scale structure of the Galactic disk, i.e., the spiral arms.
The gaps between the different arms are quite visible.

In Fig.~\ref{xydist}, we plot the spatial distribution of the molecular clouds in the $X$-$Y$
plane. The distributions of masers (Table~2 of \citealt{Xu2018b})
and young OB stars and candidates \citep{Chen2019b} are over-plotted in the diagram.
\citet{Chen2019b} have estimated  structure parameters of the Scutum, Sagittarius, Local and Perseus Arms
based on their sample of O and early B-type stars and the maser sample of \citet{Xu2018b}.
Overall, the molecular clouds catalogued here are very likely
to be spatially associated with the arms delineated by \citet{Chen2019b}.

The Perseus Arm is well traced by the molecular clouds, the masers and the OB stars in the outer
disk, for longitudes between 90\degr\ and 230\degr. The Perseus Arm is quite a diffuse arm.
Based on the spatial distribution of the OB stars  and masers, \citet{Chen2019b} identify two
possible `hole' patterns within the Arm with no masers and few OB stars. The molecular
clouds identified here around the two `holes' are also fragmented, albeit with a different pattern.
A significant fraction of the clouds
concentrated in the direction of the Galactic anti-centre ($l$ $\sim$ 180\degr), where
the masers and  OB stars also clump. Although several molecular clouds are visible in the two `holes' repoted by
\citet{Chen2019b},  the `holes' are less populated by the clouds  than in other areas.

Two significant cavities are visible in the Solar neighborhood.
Cavity~1 \citep{Chen2014, Lallement2019} falls in the direction $l$ $\sim$ 240\degr\ and
Cavity~2, which is smaller, in $l$ $\sim$ 60\degr. The model line of the Local arm,
derived from the OB stars and masers, goes along the upper boundary of
Cavity~2, and through the centre of Cavity~1. The lower boundaries of the two cavities seem to be the foreground
structure of the spur identified by \citet{Xu2016} that connects the Local and the Sagittarius Arms at
a pitch angle of $\sim$ 18\degr\ (the blue dashed line in Fig.~\ref{xydist}). On the other hand, those
nearby molecular clouds around $l$ $\sim$ 45\degr\ seem also to be the foreground
clouds of the so-called `split' \citep{Lallement2019} that connects the Local and the Sagittarius Arms at
a pitch angle of $\sim$ 48\degr\ (the pink dashed line in Fig.~\ref{xydist}).

Due to the significant amount of foreground dust extinction, we have not been able to identify the molecular clouds
of the Sagittarius Arm in the direction of~$l$~between 30\degr\ and 45\degr. The molecular clouds located
in the directions of $l$ $\sim$ 15\degr\ ($X$ $\sim$ $-$6.8\,kpc and $Y$ $\sim$ 0.5\,kpc) and $l$ $\sim$ 285\degr\
($X$ $\sim$ $-$7.5\,kpc and $Y$ $\sim$ $-$2.5\,kpc) seems to be parts of the Sagittarius Arm. Similar to the distribution of
OB stars, no molecular clouds are found in the `hole' of the  Sagittarius Arm identified by
\citet{Chen2019b}. Another significant feature is a structure of $\sim$ 3\,kpc length that connects the
the Local and the Sagittarius Arms at
a pitch angle of $\sim$ 34\degr\ (the green dashed line in Fig.~\ref{xydist}). 
This structure is apparent in the previous 3D extinction maps such as \citet{Chen2019} and \citet{Lallement2019}. 
It was treated as a cloud complex  and was labelled as ``Lower Sagittarius-Carina'' in \citet{Lallement2019}.  
However, it could be a new spur of the Galactic spiral structure and we call it ``Lower Sagittarius-Carina Spur''.
\citet{Chen2019c} have reported a massive star-forming region G352.630-1.067 and suggest that it may be
located in a spur extending from the Sagittarius Arm close to the direction of Galactic centre. The distance of G352.630-1.067
is 0.69\,kpc. It is much closer than the new spur found in the current work which have a distance of about 1.1\,kpc in the
direction. A detail analysis of this new spur will be
presented in a future work (Li et al. 2020, in preparation).

\section{Conclusion}

In this paper, we have presented a new, large and homogeneous catalogue of 567 molecular clouds within 4\,kpc from the Sun at low
Galactic latitudes ($|b|$ $<$ 10\degr) using the data presented in \ppi. The molecular clouds are identified
by a dendrogram analysis of the 3D colour excess maps of the Galactic disk. Based on the previous determinations of
extinction values and distances of over 32 million stars, we have derived accurate distances of
those molecular clouds by analyzing the colour excess and distance relations along
the sightlines of the individual clouds. The typical errors
of the distances are less than 5\,pre\,cent. As far as we know, the resulted catalogue is
the first large catalogue of molecular clouds with directly-measured
distances
in the Galactic plane.

We have also measured the areas,  the linear equivalent radii, the masses
and the surface mass densities of the catalogued  molecular clouds. A tight power-law correlation of index 1.96,
is found between the radii and masses.

We have explored the distribution of our clouds in the Galactic disk, and studied the connection between the dust distribution and the spiral arms.
Broadly speaking, the molecular clouds are found to be spatially associated with the Galactic spiral arm models delineated
in the previous work \citep{Chen2019b}, where on smaller scales, spur-like structures are not uncommon. In this respect,
we have identified a possible {spur}, the ``Lower Sagittarius-Carina Spur'', with a pitch angle of about 34\degr\ that  connects the Local
and the Sagittarius Arms in the fourth quadrant.

\section*{Acknowledgements}

We want to thank our anonymous referee for the helpful
comments.
This work is partially supported by National
Natural Science Foundation of China 11803029, U1531244, 11833006 and U1731308
and Yunnan University grant No.~C176220100007.
HBY is supported by NSFC grant
No.~11603002 and Beijing Normal University grant No.~310232102.
This research made use of astrodendro, a Python package to compute dendrograms
of Astronomical data (http://www.dendrograms.org/)

This work has made use of data products from the Guoshoujing Telescope (the
Large Sky Area Multi-Object Fibre Spectroscopic Telescope, LAMOST). LAMOST
is a National Major Scientific Project built by the Chinese Academy of
Sciences. Funding for the project has been provided by the National
Development and Reform Commission. LAMOST is operated and managed by the
National Astronomical Observatories, Chinese Academy of Sciences.

This work presents results from the European Space Agency (ESA) space mission Gaia. Gaia data are being processed by the Gaia Data Processing and Analysis Consortium (DPAC). Funding for the DPAC is provided by national institutions, in particular the institutions participating in the Gaia MultiLateral Agreement (MLA). The Gaia mission website is https://www.cosmos.esa.int/gaia. The Gaia archive website is https://archives.esac.esa.int/gaia.

\bibliographystyle{mn2e}
\bibliography{dustcloud}






\label{lastpage}
\end{document}